%% file: main.tex
\documentclass[11pt]{article}

\usepackage[final]{acl}

\usepackage{times}
\usepackage{latexsym}

\usepackage[T1]{fontenc}

\usepackage[utf8]{inputenc}

\usepackage{microtype}

\usepackage{inconsolata}

\usepackage{graphicx}
\usepackage{amsmath}
\usepackage{amssymb}
\usepackage{booktabs}
\usepackage{multirow}

\usepackage{pifont}
\usepackage{booktabs}
\usepackage{multirow}
\usepackage{graphicx} 
\usepackage{colortbl} 
\usepackage{amssymb}  
\definecolor{graybg}{gray}{0.9}
\usepackage{subcaption}
\usepackage{float}
\usepackage{xcolor}
\usepackage[most]{tcolorbox}
\usepackage{listings}
\usepackage{fontawesome5}
\newif\ifshowchanges
\showchangestrue
\ifshowchanges

\else

\fi

\definecolor{d_lightgrey}{RGB}{240, 240, 240}
\definecolor{d_darkgrey}{RGB}{169, 169, 169}
\definecolor{d_darkerGrey}{RGB}{105, 105, 105}
\definecolor{d_lightgreen}{RGB}{245, 250, 241}
\definecolor{d_darkgreen}{RGB}{207, 228, 186}
\definecolor{d_lightblue}{RGB}{241, 245, 250}
\definecolor{d_darkblue}{RGB}{186, 198, 230}
\definecolor{d_lightred}{RGB}{251, 241, 241}
\definecolor{d_darkred}{RGB}{236, 185, 191}

\newtcolorbox{prompt}{
  colback=d_lightgrey, 
  colframe=d_darkgrey, 
  colbacktitle=d_darkerGrey, 
  enhanced,
  boxrule=0pt,
  after skip=0cm,
  before skip=0.3cm,
  right skip=0cm,
  breakable,
  fonttitle=\small\bfseries, 
  fontupper=\small\linespread{1.25}\selectfont, 
  toprule=0pt,
  bottomrule=0pt,
  rightrule=0pt,
  leftrule=4pt,
  arc=0mm,
  skin=enhancedlast jigsaw,
  sharp corners,
  boxed title style={
    frame code={ 
    }
  }
}

\title{QiMeng-PRepair: Precise Code Repair via Edit-Aware Reward Optimization}

\author{
 \textbf{Changxin Ke\textsuperscript{1,2}}\;
 \textbf{Rui Zhang\textsuperscript{1,2}}\;
 \textbf{Jiaming Guo\textsuperscript{1,2}}\;
 \textbf{Yuanbo Wen\textsuperscript{1,2}}\;
 \textbf{Li Ding\textsuperscript{2,3}}\;
 \textbf{Shuo Wang\textsuperscript{1,2}}\;
\\
 \textbf{Xuyuan Zhu\textsuperscript{2}}\;
 \textbf{Xiong Peng\textsuperscript{2}}\;
 \textbf{Di Huang\textsuperscript{1,2}}\;
 \textbf{Zidong Du\textsuperscript{1,2}}\;
 \textbf{Xing Hu\textsuperscript{1,2}}\;
 \textbf{Qi Guo\textsuperscript{1,2}}\;
 \\
 \textbf{Yunji Chen\textsuperscript{1,2}}\thanks{Corresponding author.}
\\
 \textsuperscript{1}State Key Lab of Processors, Institute of Computing Technology, CAS
\\
 \textsuperscript{2}University of Chinese Academy of Sciences
\\
 \textsuperscript{3}Institute of Microelectronics, CAS
\\
\faGithub\ \href{https://github.com/kcxain/QiMeng-PRepair}{\texttt{Code}}
\quad
\raisebox{-0.2ex}{\includegraphics[height=1em]{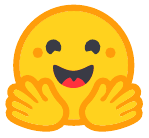}}\ \href{https://huggingface.co/collections/kcxain/qimeng-prepair}{\texttt{Models}}
\\
\\
}

\begin{document}
\maketitle
\begin{abstract}
Large Language Models (LLMs) achieve strong program repair performance but often suffer from \textit{over-editing}, where excessive modifications overwrite correct code and hinder bug localization. We systematically quantify its impact and introduce \textit{precise repair} task, which maximizes reuse of correct code while fixing only buggy parts. Building on this insight, we propose PRepair, a framework that mitigates over-editing and improves repair accuracy. PRepair has two components: \textit{Self-Breaking}, which generates diverse buggy programs via controlled bug injection and min–max sampling, and \textit{Self-Repairing}, which trains models with Edit-Aware Group Relative Policy Optimization (EA-GRPO) using an edit-aware reward to encourage minimal yet correct edits. Experiments show that PRepair improves repair precision by up to 31.4\% under $\mathrm{fix}_1@1$, a metric that jointly considers repair correctness and extent, and significantly increases decoding throughput when combined with speculative editing, demonstrating its potential for precise and practical code repair.
\end{abstract}

\input{section/1-introduction}
\input{section/2-methodology}
\input{section/3-experiments}
\input{section/4-related}
\input{section/5-conclusion}

\section*{Limitations}
Although PRepair demonstrates effective precise repair performance across multiple programming languages, it still has the following limitations:

\paragraph{Metric of Repair Precision.}
We adopt $\mathrm{fix}_p$ as the metric of repair precision, which measures edit cost as the ratio of modified lines to the total number of lines in the buggy program. While this relative formulation ensures fairness across programs of different lengths, it implicitly assumes that the perceived cost of modification scales linearly with program size. In practice, this assumption may not always hold: modifying 1 line out of 3 and modifying 10 lines out of 30 yield the same edit ratio, yet the latter arguably imposes a heavier burden on code maintainers due to the larger absolute number of changes. Designing edit cost metrics that better reflect the non-linear relationship between program length and perceived repair precision is left for future work.

\paragraph{Automatic Hyperparameter Tuning.}
Although the ablation study demonstrates the effectiveness of the accuracy threshold and penalty coefficient in EA-GRPO, the optimal settings vary across datasets with different difficulty levels. We will explore automatic tuning methods under limited computational budgets in future work.

\paragraph{Application Scope.}
PRepair focuses on function-level code repair, where LLMs are used as coding assistants to perform precise fixes. In real-world software development, bugs may appear at the file level or even the project level, where high repair precision is also required. In such scenarios, models typically generate patches directly rather than rewriting entire files or projects, which naturally aligns with the objective of minimizing edit cost. Therefore, our method remains applicable in these broader settings, and we leave a systematic investigation of its effectiveness at the file and project level as future work.

\section*{Acknowledgments}
This work is partially supported by the NSF of China (Grants No.U22A2028, 62525203, 62302483, 62341411, 6240073476), Science and Technology Major Special Program of Jiangsu (Grant No. BG2024028), Strategic Priority Research Program of the Chinese Academy of Sciences (Grants No.XDB0660300, XDB0660301, XDB0660302), CAS Project for Young Scientists in Basic Research (YSBR-029) and Youth Innovation Promotion Association CAS.


\bibliography{custom}

\clearpage
\newpage
\appendix

\input{appendix/case_study}

\input{appendix/details}
\input{appendix/additional_results}
\input{appendix/spec_edit}
\input{appendix/grpo.tex}
\input{appendix/prompts}
\end{document}

%% file: section/1-introduction.tex
\section{Introduction}
\label{sec:intro}

Program repair aims to automatically correct faulty programs while preserving their intended semantics, and has become an important research area in the era of Large Language Models~\citep{qwen25coder,surveyofapr,surveybenchmarkssolutions}. 
Prior works generally follow a structured paradigm, decomposing the task into stages such as error localization, correction, and validation~\citep{agentless,verilogcoder,Epperson_2025}. 
With the growing use of coding assistants like Copilot and Cursor, there is an increasing need for fast, end-to-end program repair models. 
To address this demand, many recent approaches employ supervised fine-tuning (SFT) and reinforcement learning (RL) to train models capable of performing program repair accurately.

\begin{figure}[t]
  \includegraphics[width=\linewidth]{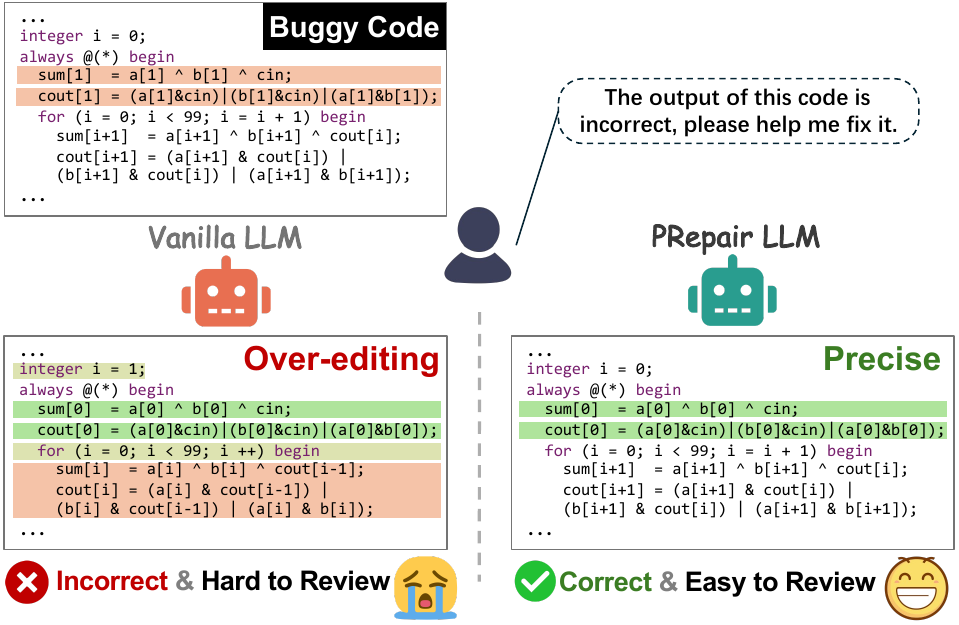}
  \caption {Existing models suffer from \textit{over-editing}, which not only reduces repair accuracy but also significantly increases the review burden for developers. In comparison, PRepair improves both repair accuracy and maintainability in practice. }
   \label{fig:1}
   \vskip -0.1in
\end{figure}


Most existing training approaches~\citep{humanevalfix,qwen25coder,morepair,slmfix} optimize repair correctness alone, treating code repair as a correctness-only objective. This formulation ignores how much the model modifies the original program. We observe that these models suffer from an \textit{over-editing} phenomenon (as illustrated in Figure~\ref{fig:1}), where they tend to regenerate large portions of the code through excessive edits instead of understanding and minimally correcting the original buggy code. Over-editing is harmful for two reasons: (1) it fails to localize the bug, thereby limiting the effectiveness of the repair; and (2) it unnecessarily rewrites the code, breaking the original structure and reducing maintainability in practice. Therefore, for code repair, precise repair is preferred, as it maximizes the reuse of correct logic in the original code while precisely fixing the buggy parts, thereby preserving code logic and reducing developers’ review burden. However, while precise repair is crucial for code repair, it remains largely unsolved in existing approaches.

Precise repair faces two key challenges: (1) Data scarcity. Effective repair requires models to understand the semantics of buggy programs, reuse their correct components, and precisely localize and fix errors. However, realistic buggy code that simultaneously contains substantial correct logic and localized faults is extremely scarce. (2) Preservation of correct code. During training, it is challenging to make the model aware of how much of the code has been edited, so that it preserves the correct parts while precisely localizing and fixing only the buggy portions.

To address the over-editing issue, we propose the \textbf{PRepair} framework, which explicitly guides models to perform precise repairs.
Our central insight is that optimizing for minimal yet sufficient edits preserves repair correctness while encouraging faithful reuse of correct program logic. 
To address the two challenges of precise repair, the PRepair framework consists of two steps: 
(1) \textit{Self-Breaking}, where we design a precise code repair data generation framework that systematically injects bugs into correct programs to construct large-scale training data, combined with a min–max sampling strategy to maximize the diversity of buggy programs while avoiding over-concentration on similar bug patterns; 
(2) \textit{Self-Repairing}, where the model is optimized with proposing Edit-Aware Group Relative Policy Optimization (EA-GRPO) to encourage both correct and minimal code repairs. EA-GRPO introduces an edit-aware reward, where edit penalties are applied when the model achieves sufficient repair correctness. This design effectively balances repair correctness and extent, encouraging minimal yet accurate code fixes. 
Besides, to evaluate both repair correctness and the extent of modifications, we introduce $\mathrm{fix}_p@k$, the first metric specifically designed for assessing precise repair, which jointly considers correctness and the number of edits.


Compared with previous methods that optimize code repair solely for correctness, our method offers two key advantages. 
First, the model learns to focus its attention on the buggy lines, acquiring an implicit error localization ability that guides precise repair, which not only improves repair accuracy but also enhances cross-domain code repair capability.
Second, by following the logic of the buggy code, it reuses correct portions of the original program, alleviating the over-editing problem and improving maintainability in practice, as shown in Figure~\ref{fig:1}. 

Experiments on two models of different sizes and two fundamentally different languages, Python and Verilog, show that PRepair effectively reduces unnecessary edits while improving repair correctness. In addition, when combined with speculative editing, PRepair enables faster inference, demonstrating its practical value and generality across diverse programming languages. The main contributions of this paper are as follows:



\begin{itemize}
\item We identify over-editing as a key issue in LLM-based code repair under GRPO and propose $\mathrm{fix}_p@k$, the first metric for evaluating repair precision.
\item We propose the PRepair framework to enhance code repair without labeled data, and introduce EA-GRPO to train models for precise code repair using an edit-aware reward.
\item Empirical evaluations on multiple models and benchmarks demonstrate that PRepair achieves superior repair precision and correctness.
\item When combined with speculative editing, EA-GRPO significantly increases inference throughput, highlighting the practical value of PRepair as real-world code assistance.
\end{itemize}

%% file: section/2-methodology.tex
\section{Methodology}
\label{sec:method}

In this section, we first analyze existing models trained with naive GRPO and empirically study the relationship between repair accuracy and extent of modifications. Motivated by these findings, we introduce a novel metric, $\mathrm{fix}_p$, which jointly measures repair accuracy and the number of edited lines. Based on this metric, we then present the proposed \textbf{PRepair} framework.

\begin{figure}[t]
    \centering
    \begin{subfigure}[t]{\linewidth}
        \centering
        \includegraphics[width=0.95\linewidth]{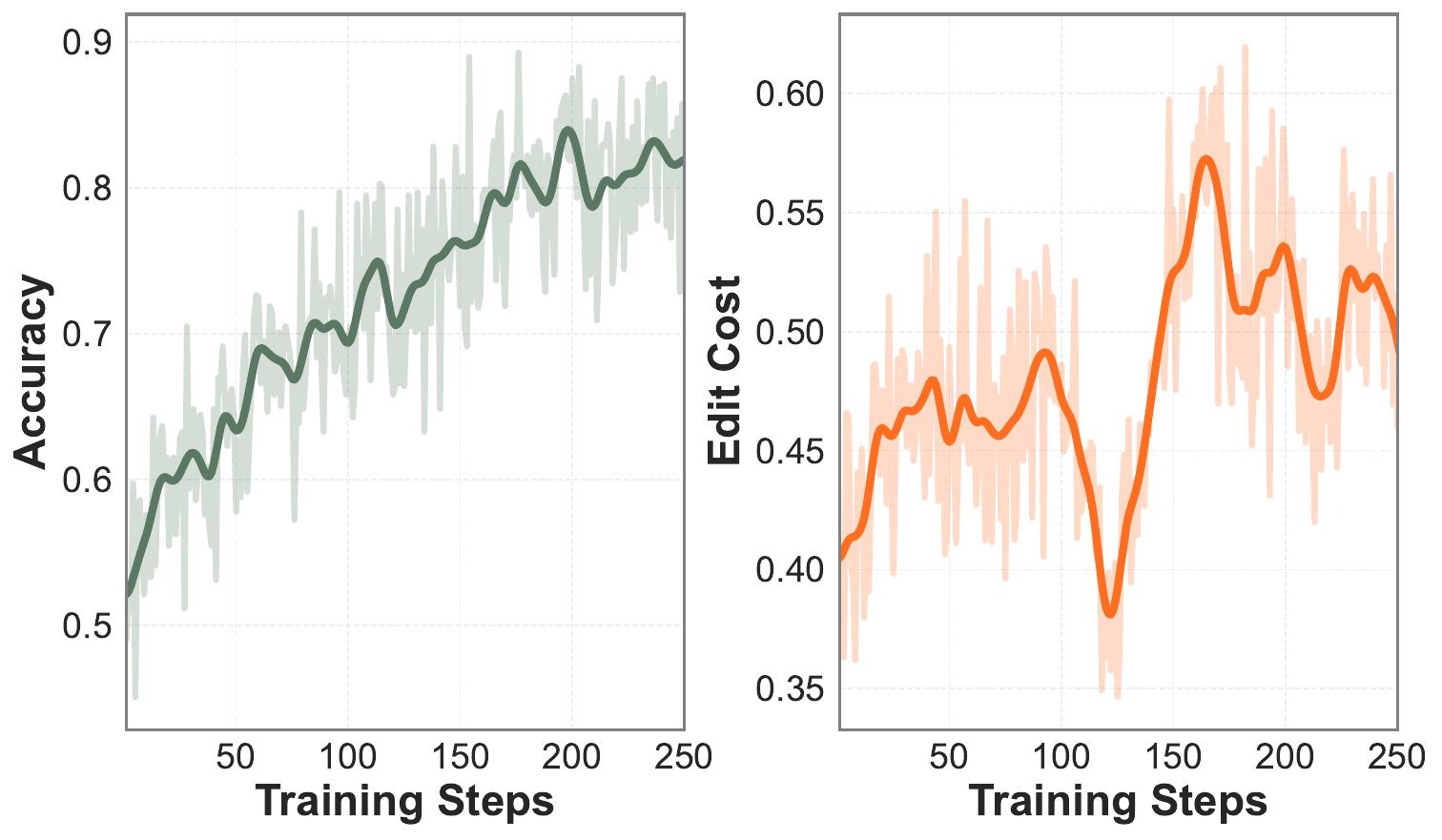}
        \caption{Python Code Repair}
        \label{fig:sub1}
    \end{subfigure}

    \vspace{0.6em}

    \begin{subfigure}[t]{\linewidth}
        \centering
        \includegraphics[width=0.95\linewidth]{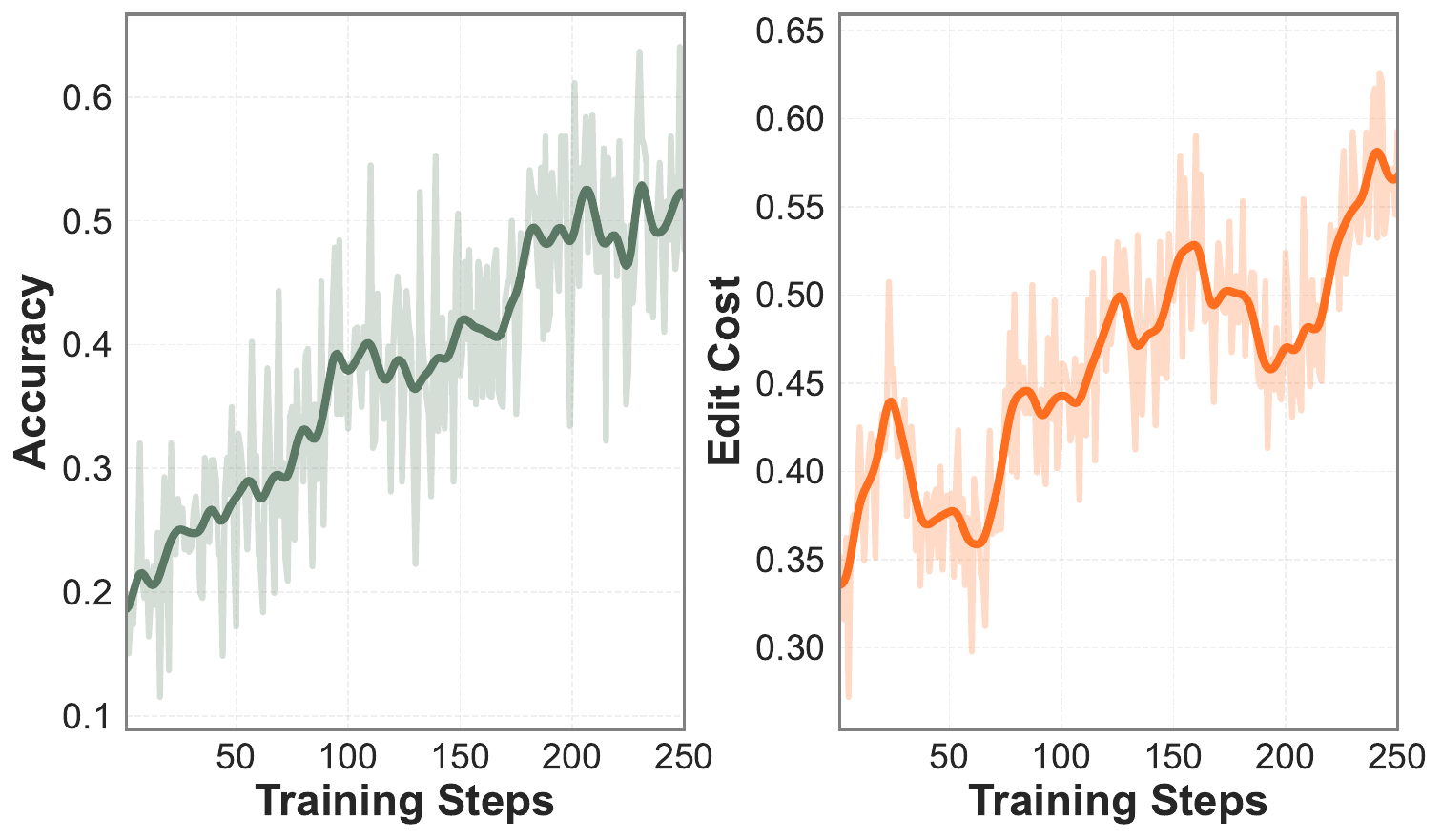}
        \caption{Verilog Code Repair}
        \label{fig:sub2}
    \end{subfigure}

    \caption{\textbf{GRPO training with correctness-only rewards.} For both Python and Verilog code repair tasks, although performance improves during training, the edit cost increases substantially, leading to a more severe over-editing phenomenon.}
    \label{fig:observation}
    \vskip -0.15in
\end{figure}

We model code as a sequence of lines $X = \{x_1, x_2, \dots, x_n\}$. Given a buggy program, the goal of program repair is to perform the necessary line-level insertions, deletions, and substitutions to produce a corrected sequence $Y = \{y_1, y_2, \dots, y_m\}$ that satisfies the intended functionality. To quantify the distance between the buggy code and the corrected code, we introduce the Edit Cost $\mathbf{D}_\mathrm{EC}$, which is based on the Levenshtein distance $\mathbf{D}(X, Y)$~\cite{Levenshtein}. This distance measures the minimum number of insertions, deletions, and substitutions required to transform one code into the other. Let $|X|$ denote the number of lines in the source program. We define Edit Cost as:
\[
\mathbf{D}_\mathrm{EC}(X, Y) = \frac{\mathbf{D}(X, Y)}{|X|}
\]
Here, dividing $|X|$ normalizes the edit distance by lines of buggy code, allowing fair comparison across programs of different sizes.

\subsection{Observations}

In this section, we explore the phenomenon of \textit{over-editing} in LLMs and investigate the relationship between code repair accuracy and edit cost.

We conduct experiments on Python and Verilog code repair tasks. The Python dataset is collected from LeetCodeDataset~\citep{lcd}, and the Verilog dataset is obtained from QiMeng-CodeV-R1~\citep{codev-r1}. We design a reward that considers only repair correctness, and the model performance and edit cost are shown in the Figure \ref{fig:observation}. We find that as training progresses, repair correctness improves, but over-editing becomes increasingly severe. The model does not learn to fix errors precisely but instead makes large modifications to ``hit'' a correct solution. As training continues, the edit cost even exceeds 0.6, indicating that the model introduces extensive redundant changes without understanding the original buggy code and localizing the errors. These findings show that evaluating code repair solely based on correctness is insufficient, which motivates the need to design a metric that explicitly measures repair precision and to incorporate edit cost into training.

\subsection{Metric Design}
\label{section:metric_design}
Considering the over-editing phenomenon, to better capture precise code repair capability, 
we propose $\mathrm{fix}_p@k$, a novel metric that jointly accounts for repair correctness and edit cost.
To reduce statistical bias, we adopt an unbiased estimation method by sampling $n$ candidates. 
The computation of a general metric $(\cdot)@k$ is defined as:
\begin{equation*}
(\cdot)@k = 1 - \frac{{n - c \choose k}}{{n \choose k}},
\end{equation*}
where $c$ denotes the number of samples that satisfy the corresponding \textit{checking criterion} among the $n$ generated candidates and $k$ represents the number of candidates considered.


Given the golden fixed program $Y$ and the model-generated fix $Y'$,  
We define ${\text{fix}_p@k}$, where $p$ denotes the ratio between the acceptable  
edit cost and the theoretical minimum Edit Cost, representing the tolerance level for repair cost in evaluation. The corresponding checking criterion is:
\[
c = \sum_{i=1}^{n} 
\mathbb{I} \left[
    \mathrm{correct}_i \;\land\;
    \left( \frac{\mathbf{D}_\mathrm{EC}(X, Y')}{\mathbf{D}_\mathrm{EC}(X,Y)} \le p \right)
\right].
\]
Here, $\mathrm{correct}_i$ indicates that the $i$-th generated code passes all tests. We also report the correctness only metric using pass@k~\citep{humaneval}.

\begin{figure*}[t]
  \includegraphics[width=\linewidth]{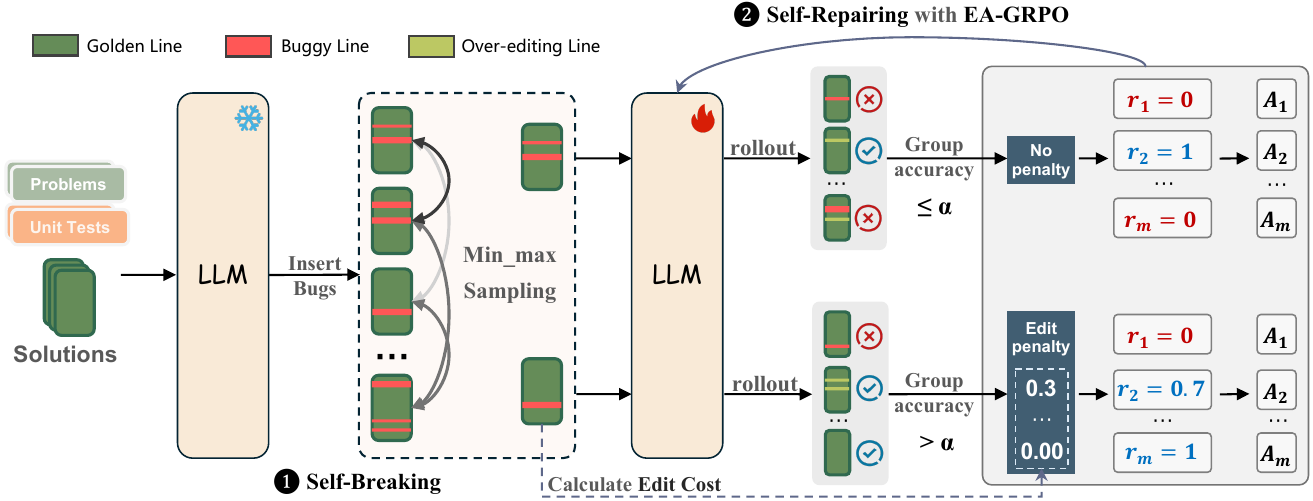}
  \caption {\textbf{Overview of the PRepair framework.} It consists of two stages: Self-Breaking, where the model injects diverse bugs into golden programs to construct high-quality buggy inputs, and Self-Repairing, where the model learns to precise repair these buggy programs via EA-GRPO which uses a dynamic edit-aware reward to encourage minimal yet correct edits.}
  \label{fig:framework}
  \vskip -0.1in
\end{figure*}

\subsection{PRepair framework}
Program repair is challenged by the lack of realistic buggy data with localized faults and by the difficulty of preserving correct code during repair. To address this challenge, we propose the PRepair framework, as shown in Figure~\ref{fig:framework}, which consists of two stages: (1) Self-Breaking, where the model generates high-quality buggy code by itself without human annotations. (2) Self-Repairing, where the model is trained with EA-GRPO to improve its ability of precise code repair. 


\paragraph{Self-Breaking.}
Given a program description and its corresponding golden code $Y$, we prompt the model to inject bugs (detailed prompt is in Appendix~\ref{appendix:details}) into $Y$ and sample a set of buggy programs $\mathcal{X} = \{X_1, X_2, \dots, X_m\}$. 
To improve computational efficiency while preserving bug diversity, we adopt a min-max sampling strategy. Specifically, we select a subset $\mathcal{X}_s \subset \mathcal{X}$ by minimizing the maximum pairwise similarity among buggy samples, where similarity is defined as $1 - \mathbf{D}_\mathrm{EC}(X_i, X_j)$. The selected subset is obtained by solving:
\[
\mathcal{X}_s
=
\min_{\substack{
\mathcal{X}' \subset \mathcal{X} \\
|\mathcal{X}'| = k
}}
\;
\max_{\substack{
X_i, X_j \in \mathcal{X}' \\
i \neq j
}}
\;
\big( 1 - \mathbf{D}_\mathrm{EC}(X_i, X_j) \big).
\]
This strategy encourages the sampled buggy programs to be maximally diverse in terms of edit distance, resulting in a more diverse and informative set of buggy programs for training.

\paragraph{Self-Repairing.} 
Given a program description and its corresponding buggy code set $\mathcal{X}_s$ sampled from the Self-Breaking stage, the objective of this stage is to train the model to repair the buggy programs and improve its repair policy. Specifically, the model generates candidate repairs for each buggy input, and the policy is updated using the proposed \textit{Edit-Aware Group Relative Policy Optimization (EA-GRPO)}. During optimization, rewards are computed with a dynamic edit-aware reward, which jointly considers repair correctness and edit cost to guide the model toward accurate and minimal code fixes.

\subsection{EA-GRPO}

Program repair differs from code generation. Using a binary reward based solely on correctness is insufficient because it cannot reflect the model's ability to precisely identify errors.

To address this, we design the EA-GRPO mechanism that encourages minimal and precise changes while ensuring correctness. Specifically, to avoid over-penalizing model edits that could harm correctness, the penalty in EA-GRPO is applied dynamically. It is triggered only when the model achieves sufficient group-level accuracy.
Compared with naive GRPO~\cite{grpo} (details can be found at Appendix~\ref{detals_grpo}), EA-GRPO introduces a dynamic edit-aware reward, focusing on balancing repair correctness and edit cost.

\paragraph{Group Accuracy Threshold.}
During training, given a buggy input $X_t\in \mathcal{X}_s$, we compute the average repair accuracy $\mathrm{Acc}_{\mathcal{G}^t}$ of its rollout group $\mathcal{G}^t = \{o_1, o_2, \dots, o_n\}$, where each $o_i$ denotes a repaired code generated by the model. The edit penalty is activated only when the group-level accuracy exceeds a threshold $\alpha$, formally defined as
\[
\mathcal{T}(\mathcal{G}^t) =
\begin{cases}
1, & \text{if } \mathrm{Acc}_{\mathcal{G}^t} \ge \alpha, \\
0, & \text{otherwise}.
\end{cases}
\]

\paragraph{Dynamic Edit-Aware Reward Shaping.}
For correct samples in groups that satisfy the accuracy threshold, we apply a standardized edit penalty to encourage correct repairs with minimal edit cost. Let $\mathcal{G}_c^t \subset \mathcal{G}^t$ denote the set of correct samples. The penalty for each sample $o_i \in \mathcal{G}_c^t$ is defined as
\[
\mathcal{P}^{\mathcal{G}}_i
=
\sigma\!\left(
\frac{
\mathbf{D}_\mathrm{EC}(X_t, o_i)
-
\mathrm{mean}(\mathbf{D}_\mathrm{EC}(X, \mathcal{G}_c^t))
}{
\mathrm{std}(\mathbf{D}_\mathrm{EC}(X, \mathcal{G}_c^t))
}
\right),
\]
where $\mathrm{mean}(\mathbf{D}_\mathrm{EC}(X, \mathcal{G}_c^t))$ and $\mathrm{std}(\mathbf{D}_\mathrm{EC}(X, \mathcal{G}_c^t))$ are the mean and standard deviation of the edit cost for correct samples in the group. The outer sigmoid bounds the penalty while preserving the relative ordering of edit costs within the group.

\paragraph{Reward Design.}
The reward for each sample in the group is then defined as
\[
\mathcal{R}^{\mathcal{G}}_i
=
\begin{cases}
1 - \mathcal{T}(\mathcal{G}) \cdot \beta \cdot \mathcal{P}^{\mathcal{G}}_i, & \text{if } o_i \text{ is correct}, \\
0, & \text{if } o_i \text{ is incorrect}.
\end{cases}
\]
where $\beta$ is a penalty coefficient controlling the strength of the edit penalty. Importantly, the computation of this reward function does not require the golden code, it only uses the edit cost between the buggy input $X$ and the generated samples.

\subsection{Speculative Edits}

Speculative decoding~\citep{spec_start} is widely used in code editing scenarios because the original code can be reused across successive edits. We adopt Prompt Lookup Decoding~\citep{spec_edit}, a speculative decoding method, to accelerate inference. Speculative decoding improves generation efficiency by first producing multiple draft tokens and then verifying them in parallel. Unlike conventional approaches that rely on a separate draft model, Prompt Lookup Decoding directly reuses the input prompt as the draft through $n$-gram matching, which is particularly well suited for code editing scenarios. For this reason, it is also referred to as \emph{Speculative Edits}. Our work focuses on reducing the edit cost between the input buggy code and the output, which substantially increases the acceptance rate of speculative edits. A detailed theoretical derivation is provided in Appendix~\ref{spec_edit}. Given a speculative window of $K$ draft tokens per decoding step, the decoding throughput $T$ (tokens/s) can be expressed as
\[
T \propto \frac{1 - (1 - \mathbf{D}_\mathrm{EC})^{K+1}}{\mathbf{D}_\mathrm{EC}}.
\]
It shows that reducing the edit cost leads to a significant increase in throughput. Therefore, when applying speculative edits, a smaller edit cost directly translates to a larger speedup.

%% file: section/3-experiments.tex
\section{Experiment}
\label{sec:experiments}

\subsection{Experimental Setup}

\subsubsection{Benchmarks}
We form a code repair benchmark that spans multiple programming languages and paradigms and covers diverse real-world errors, enabling a comprehensive evaluation of model code repair capabilities. The statistics of the two benchmarks are shown in Table \ref{tab:benchmark_stats}. 

\paragraph{Python code repair.} We follow HumanEvalFix~\citep{humanevalfix}, which extends the original HumanEval benchmark. It provides buggy code functions with subtle errors and corresponding unit tests, and models are tasked with generating correct fixes. Bugs are manually introduced to original HumanEval solutions so that the code still runs but fails at least one test. The benchmark covers various types of logical errors, including missing logic, excess logic, and wrong logic such as value, operator, variable, or function misuse, totaling 164 buggy samples.

\paragraph{Verilog code repair.} Existing Verilog code repair benchmarks~\citep{rtlfixer} have clear limitations. Most of them mainly target syntax errors and give little attention to functional errors. Our work aims to enable LLMs to reuse correct logic in buggy code and apply precise and minimal fixes. We systematically summarize common logical error patterns in Verilog from~\citet{rtlfixer,hdldebugger,hdlroot} and prompt models to inject these bugs into correct code from the QiMeng-CodeV-R1~\citep{codev-r1} dataset. This process produces a diverse Verilog code repair benchmark with 352 samples.

\begin{table*}[t]
\centering
\resizebox{\textwidth}{!}{
\begin{tabular}{llcccccccccccc}
\toprule
\multirow{2}{*}{Language} & \multirow{2}{*}{Method}
& \multicolumn{3}{c}{\textbf{$\mathrm{pass}@k$}} 
& \multicolumn{3}{c}{\textbf{$\mathrm{fix}_{1}\!@\!k$}} 
& \multicolumn{3}{c}{\textbf{$\mathrm{fix}_{1.5}\!@\!k$}} 
& \multicolumn{3}{c}{\textbf{$\mathrm{fix}_{2}\!@\!k$}} \\

\cmidrule(lr){3-5} \cmidrule(lr){6-8} \cmidrule(lr){9-11} \cmidrule(lr){12-14}
 & & 1 & 5 & 10 & 1 & 5 & 10 & 1 & 5 & 10 & 1 & 5 & 10 \\
\midrule
\multirow{12}{*}{\textbf{Python}}
 & \textbf{GPT4} & 84.51 & 94.68 & 96.95 & 52.20 & 71.50 & 76.83 & 53.29 & 72.81 & 78.66 & 60.73 & 79.85 & 84.76 \\
 & + Prompt & 86.89 & 95.90 & 96.95 & 62.56 & 80.28 & 85.98 & 63.90 & 81.36 & 86.59 & 72.44 & 87.72 & 90.24 \\
\cmidrule(lr){2-14}
 & \textbf{Gemini2.0-flash} & 90.73 & 91.94 & 92.07 & 44.88 & 47.71 & 48.78 & 47.38 & 49.98 & 51.22 & 55.12 & 58.60 & 60.37 \\
 & + Prompt & 92.20 & 93.28 & 93.29 & 63.78 & 66.45 & 67.07 & 64.70 & 67.67 & 68.29 & 71.71 & 74.95 & 75.61 \\
\cmidrule(lr){2-14}
 & \textbf{Qwen2.5-Coder-3B} & 53.78 & \underline{82.22} & \underline{86.48} & 33.72 & 55.91 & 60.93 & 34.18 & 56.68 & 62.16 & \underline{38.78} & 63.99 & 69.74 \\
 & + Prompt & 50.91 & 57.34 & 63.18 & 32.13 & \underline{57.34} & \underline{63.18} & 32.32 & \underline{57.74} & \underline{63.73} & 36.83 & \underline{65.44} & \underline{71.92} \\
 & + GRPO & \textbf{80.52} & \textbf{87.26} & \textbf{88.43} & \underline{34.27} & 40.37 & 41.66 & \underline{36.01} & 41.67 & 42.89 & \underline{45.61} & 52.71 & 54.67 \\
 \rowcolor{blue!10} \cellcolor{white} & + EA-GRPO & \underline{79.05} & 80.97 & 81.09 & \textbf{67.96} & \textbf{70.67} & \textbf{71.03} & \textbf{67.96} & \textbf{70.67} & \textbf{71.03} & \textbf{74.36} & \textbf{77.22} & \textbf{77.43} \\
 \cmidrule(lr){2-14}
 & \textbf{Qwen2.5-Coder-7B} 
 & 86.28 & \underline{92.91} & \textbf{93.79} & 60.67 & 72.37 & 74.24 & 61.59 & 72.45 & 74.25 & 71.46 & 81.69 & 83.27 \\
 & + Prompt 
 & 87.23 & \textbf{92.37} & \underline{93.23} & \underline{66.52} & \underline{76.44} & \underline{78.15} & \underline{66.86} & \underline{76.51} & \underline{78.15} & \underline{77.41} & \underline{85.58} & \underline{86.78} \\
 & + GRPO 
 & \underline{89.82} & 91.70 & 91.93 & 47.44 & 48.91 & 49.09 & 48.20 & 50.02 & 50.29 & 60.88 & 62.31 & 62.50 \\
 \rowcolor{blue!10} \cellcolor{white} & + EA-GRPO 
 & \textbf{91.19} & 92.90 & 93.22 & \textbf{81.62} & \textbf{83.51} & \textbf{84.01} & \textbf{82.13} & \textbf{83.76} & \textbf{84.08} & \textbf{89.54} & \textbf{91.61} & \textbf{92.00} \\
\midrule

\multirow{12}{*}{\textbf{Verilog}}
 & \textbf{GPT4} & 69.52 & 85.44 & 89.49 & 2.30 & 4.69 & 5.97 & 3.84 & 7.76 & 9.38 & 5.77 & 11.65 & 14.20 \\
 & + Prompt & 55.99 & 79.83 & 84.38 & 22.13 & 44.84 & 52.56 & 28.04 & 53.49 & 61.65 & 33.92 & 58.85 & 65.34 \\
\cmidrule(lr){2-14}
 & \textbf{Gemini2.0-flash} & 56.65 & 69.05 & 72.44 & 19.06 & 23.25 & 24.43 & 24.01 & 30.44 & 32.39 & 30.57 & 38.15 & 40.06 \\
 & + Prompt & 68.01 & 74.46 & 76.99 & 42.33 & 48.09 & 50.00 & 48.44 & 54.34 & 56.25 & 53.49 & 59.02 & 61.08 \\
\cmidrule(lr){2-14}
& \textbf{Qwen2.5-Coder-3B} 
& 45.91 & 59.93 & 63.57 & 34.08 & 50.06 & \underline{54.57} & 36.14 & \underline{52.14} & \underline{56.50} & \underline{39.91} & \underline{55.63} & \underline{59.66} \\
& + Prompt  
& 44.53 & 57.55 & 60.52 & \underline{34.20} & \underline{50.54} & 54.54 & \underline{36.18} & 51.90 & 55.57 & 39.52 & 54.65 & 57.96 \\
& + GRPO    
& \underline{47.90} & \underline{61.08} & \underline{65.44} & 18.55 & 33.64 & 39.34 & 21.52 & 37.41 & 43.51 & 28.65 & 44.52 & 50.10 \\
\rowcolor{blue!10} \cellcolor{white} & + EA-GRPO & \textbf{52.64} & \textbf{66.49} & \textbf{69.93} & \textbf{37.40} & \textbf{54.03} & \textbf{58.15} & \textbf{40.80} & \textbf{56.79} & \textbf{60.83} & \textbf{45.30} & \textbf{60.29} & \textbf{63.69} \\
\cmidrule(lr){2-14}
& \textbf{Qwen2.5-Coder-7B} 
& 57.36 & 69.31 & 72.67 & 36.70 & 50.29 & 54.31 & 42.86 & \underline{55.85} & \underline{59.29} & 48.98 & \underline{61.47} & \underline{64.34} \\
& + Prompt  
& 57.07 & 68.63 & 72.10 & \underline{38.00} & \underline{51.10} & \underline{54.74} & \underline{43.81} & 55.75 & 58.62 & \underline{49.59} & 61.25 & 64.14 \\
& + GRPO    
& \underline{68.37} & \underline{71.91} & \textbf{72.89} & 8.49 & 9.95 & 10.21 & 12.93 & 15.69 & 16.24 & 23.85 & 27.29 & 28.04 \\
\rowcolor{blue!10} \cellcolor{white} & + EA-GRPO & \textbf{68.66} & \textbf{72.02} & \underline{72.75} & \textbf{68.11} & \textbf{71.38} & \textbf{72.07} & \textbf{68.11} & \textbf{71.38} & \textbf{72.07} & \textbf{68.59} & \textbf{71.85} & \textbf{72.61} \\
\bottomrule
\end{tabular}
}
\caption{\textbf{Main results.}
We report $\mathrm{pass}@k$ and $\mathrm{fix}_p@k$ results with $k \in \{1, 5, 10\}$ and $p \in \{1, 1.5, 2\}$. 
We evaluate GPT4 and Gemini2.0-flash with prompt engineering, as well as Qwen2.5-Coder-3B and Qwen2.5-Coder-7B under prompt engineering, GRPO, and our EA-GRPO. \textbf{Bold} indicates the best result, and \underline{underline} indicates the second best in the same model.}
\vskip -0.15in
\label{tab:main_results}
\end{table*}

\begin{figure*}[t]
    \centering
    \includegraphics[width=\textwidth]{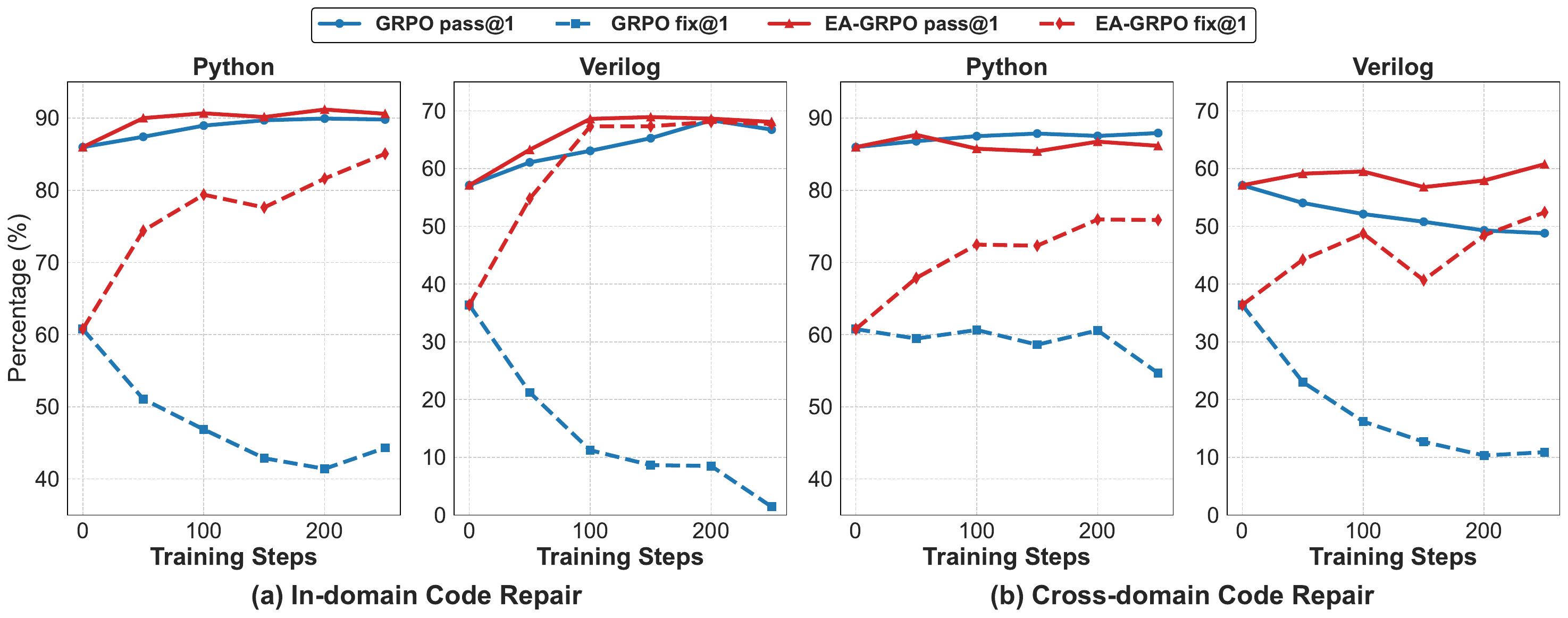}
    \vskip -0.05in
    \caption{\textbf{Code repair performance of in-domain and cross-domain.} We plot the changes of $\mathrm{pass}@1$ and $\mathrm{fix}_1@1$ across training steps, reporting both in-domain and cross-domain performance. (a) In-domain: models are trained on Python data and evaluated on Python code repair; similarly for Verilog. (b) Cross-domain: models trained on Python data are evaluated on Verilog code repair, and vice versa.}
    \label{fig:main_results_comparison}
    \vskip -0.15in
\end{figure*}

\begin{figure}[t]
    \includegraphics[width=\linewidth]{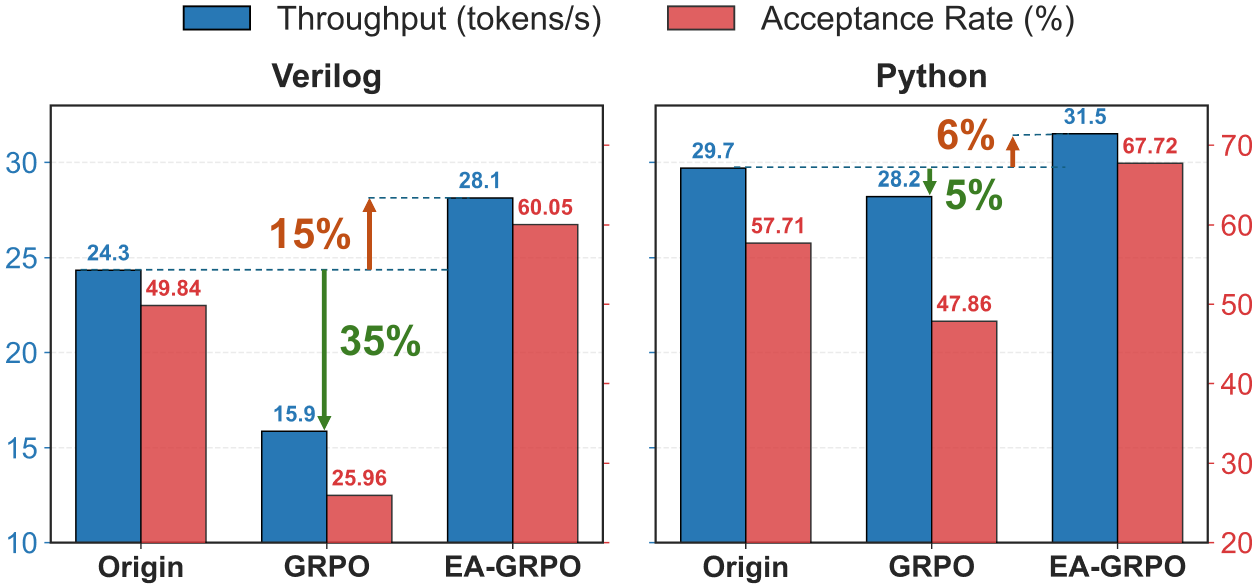}
    
    \caption{\textbf{Decoding Performance with Speculative Edits.} Throughput and acceptance rates of Origin (before training), GRPO, and EA-GRPO on Python and Verilog benchmarks, using buggy code as draft.}
    \label{fig:spec_edit}
    \vskip -0.15in
\end{figure}

\subsubsection{Base model \& Baselines}

\paragraph{Models.} 
We conduct experiments on Qwen2.5-Coder-3B and Qwen2.5-Coder-7B~\citep{qwen25coder}, two models of different scales, to evaluate the generality of our approach across model capacities.

\paragraph{Baselines.}
We compare our approach with several baselines. (1) \textbf{Prompt Engineering} instructs the model to perform code repair with minimal modifications via prompts. Specifically, we append the instruction ``Please make sure to make minimal changes to the buggy code.'' at the end of the prompt. (2) \textbf{GRPO} performs reinforcement learning using the same training data, number of training steps, and hyperparameters as EA-GRPO. The only difference is that its reward function considers repair correctness only, without incorporating any edit-aware terms. (3) In addition, we evaluate two widely used commercial code assistant models, GPT4~\citep{gpt4} and Gemini2.0-flash~\citep{gemini}. For these strong proprietary models, we apply the same prompt engineering strategy to assess how much prompt-based guidance alone can improve repair precision.

\subsubsection{Implementation Details}

For Python code repair, we use the training data from \cite{lcd}, which contains 2,869 Python programming tasks crawled from LeetCode, each equipped with comprehensive test suites. In the Self-Breaking stage, we first prompt the model to sample $|\mathcal{X}|=32$ buggy variants for each task, and then apply a min-max sampling strategy to reduce the number of samples to $|\mathcal{X}'|=4$. We further filter out false buggy cases that still pass all test cases. This process results in a final dataset of 10,242 <program description, buggy code> pairs. For Verilog code repair, we use the training data from QiMeng-CodeV-R1~\citep{codev-r1}, which contains 3,033 Verilog programming tasks, each provided with golden reference code and rule-based verification tools to validate the correctness of generated programs. Using the same parameters as in Python code repair, the Self-Breaking step yields 11,200 buggy code samples.

We conduct reinforcement learning training using the VeRL framework~\citep{verl}. More details and training hyperparameters are provided in Appendix~\ref{appendix:details}.

\subsection{Results and Analysis}

Our main results and comparisons with the baselines are presented in Table \ref{tab:main_results} and Figure \ref{fig:main_results_comparison}.

\subsubsection{Main Results}

\paragraph{Training is Necessary.}
As shown in Table~\ref{tab:main_results}, we report the results of applying prompt engineering to GPT4, Gemini2.0-Flash, Qwen2.5-Coder-3B, and Qwen2.5-Coder-7B. Our results reveal that prompt engineering introduces substantial uncertainty in model behavior. For Python code repair, this strategy has little impact on $\text{pass}@k$ and yields only limited improvements in $\mathrm{fix}_p@k$. In contrast, for Verilog code repair, prompt engineering significantly degrades performance of GPT4, reducing $\text{pass}@1$ by 13.53\%. These observations indicate that prompt engineering is far less effective than EA-GRPO. GPT-4 and Gemini 2.0 Flash achieve substantially lower $\mathrm{fix}_1@1$ than Qwen2.5-Coder-7B trained with EA-GRPO. On Python, their $\mathrm{fix}_1@1$ is lower by 19.10\% and 17.84\%, respectively. On Verilog, the gap is even larger, with drops of 45.98\% and 25.78\%. These results show that training with EA-GRPO is necessary.

\paragraph{Fewer Edits, More Correct Repairs.}
As shown in Table~\ref{tab:main_results}, under the $\mathrm{fix}_p$ metric, EA-GRPO substantially improves repair precision on both languages. Specifically, $\mathrm{fix}_1@1$ increases by 20.95\% on Python and by 31.41\% on Verilog compared to the original model, significantly alleviating the over-editing phenomenon. In contrast, models trained with GRPO exhibit a severe degradation in $\mathrm{fix}_p$. On Verilog, $\mathrm{fix}_1@1$ drops sharply from 36.70\% to 8.49\%, and $\mathrm{fix}_2@1$ decreases from 48.98\% to 23.85\%, reflecting pronounced over-editing behavior that substantially increases the code review burden for developers.


Notably, EA-GRPO also yields consistent gains in repair correctness in most settings. Compared with GRPO, Qwen2.5-Coder-7B trained with EA-GRPO improves pass@1 by 1.37\% on Python and by 0.29\% on Verilog, while Qwen2.5-Coder-3B achieves a 4.74\% improvement in pass@1 on Verilog. These results indicate that explicitly encouraging fewer edits does not harm repair correctness; instead, it helps the model better understand the original program logic and more accurately localize bugs, leading to more effective repairs. In a small number of cases, such as Qwen2.5-Coder-3B on Python, pass@1 is slightly lower than that of GRPO (by 1.47\%). However, this minor drop is accompanied by a substantial improvement in $\mathrm{fix}_p@k$, demonstrating that EA-GRPO successfully balances repair correctness and edit cost.

We further present a case study in Appendix \ref{appendix:case_study} and Figure \ref{fig:heat_map}. The results show that models trained with EA-GRPO generate fixes that better follow the logic of the buggy code, while placing substantially higher attention on the buggy lines. This indicates that the model learns to reuse the correct parts of the original program and precisely localize and repair the buggy components.

\begin{table}[t]
\centering
\resizebox{\linewidth}{!}{
\begin{tabular}{c c c c c c c}
\toprule
$\alpha$ & $\beta$ & pass@1 & pass@5 & fix$_1$@1 & fix$_{1.5}$@1 & fix$_2$@1 \\
\midrule
0   & 0.05 & \underline{90.73} & \underline{92.66} & \textbf{81.74} & \textbf{82.35} & \underline{88.93} \\
0.5 & 0.2  & 85.06 & 85.36 & 80.27 & 80.27 & 84.09 \\
0.8 & 0.05 & \textbf{91.19} & \textbf{92.90} & \underline{81.62} & \underline{82.13} & \textbf{89.54} \\
0.8 & 0.25 & 88.78 & 91.22 & 77.93 & 78.54 & 86.62 \\
1.1 & /    & 89.82 & 91.70 & 47.44 & 48.20 & 60.88 \\
\bottomrule
\end{tabular}
}
\caption{Ablation results of EA-GRPO on Qwen2.5-Coder-7B for Python code repair with varying $\alpha$ and $\beta$, reporting pass@1, pass@5, and fix$_p$@1 metrics.}
\label{tab:abaltion}
\vskip -0.15in
\end{table}

\paragraph{Better Cross-domain Generalization.}
We evaluate cross-domain generalization by assessing the Verilog code repair performance of models trained on Python and, conversely, the Python code repair performance of models trained on Verilog.
As presented in Figure \ref{fig:main_results_comparison}. We observe that in cross-domain settings, the models trained with EA-GRPO maintain stable repair correctness while significantly improving $\mathrm{fix}_1@1$. In contrast, the models trained with GRPO exhibits a notable drop in $\mathrm{fix}_1@1$, indicating increased edit cost, and its $\mathrm{pass@1}$ is also unstable. For instance, when trained on Python data and evaluated on Verilog code repair, $\mathrm{pass@1}$ of GRPO decreases from 57.12\% to 48.81\% (a drop of 8.31\%), and $\mathrm{fix}_1@1$ drops from 36.38\% to 10.88\% (a drop of 26.50\%). This demonstrates that optimizing solely for correctness does not enable the model to generalize its understanding of code or to accurately localize bugs. By contrast, EA-GRPO encourages the model to reuse correct portions of the buggy code while precisely localizing errors, achieving better cross-domain generalization.

\paragraph{Faster Repair via Speculative Edits}
As shown by the throughput and acceptance rate in Figure~\ref{fig:spec_edit} and Table~\ref{table:spec_edit}, EA-GRPO substantially increases the draft token acceptance rate due to its significantly reduced edit cost, resulting in up to a 15\% improvement in decoding throughput. In contrast, GRPO exacerbates over-editing, leading to a throughput degradation of up to 35\%. These results demonstrate the practical significance of our method: when deployed in real-world code assistants, EA-GRPO can markedly improve online serving efficiency while maintaining high repair quality.

\subsection{Ablation Study}

To investigate the effectiveness of EA-GRPO, we conduct an ablation study on Qwen2.5-Coder-7B as shown in Table \ref{tab:abaltion}. Specifically, we vary the Group Accuracy Threshold $\alpha$, which controls when the edit penalty is applied: $\alpha=0$ applies the penalty to all correct samples, whereas $\alpha=1.1$ uses only the correctness reward. We also experiment with different values of the penalty coefficient $\beta$. These ablations illustrate the impact of EA-GRPO on balancing repair correctness and minimal edits. In particular, increasing $\beta$ may reduce $\mathrm{pass}@1$, which in turn lowers $\mathrm{fix}@k$. On the other hand, setting $\alpha$ too low penalizes all samples, causing the model to neglect correctness, while setting $\alpha$ too high prevents the model from learning precise repairs. Both extremes can degrade performance.

%% file: section/4-related.tex
\section{Related Work}
\label{sec:related}
\paragraph{Buggy Data Construction.}
In software, benchmarks for function-level code repair differ mainly in how buggy programs are generated. QuixBugs~\citep{quixbugs} contains only 40 programs, limiting coverage. DebugBench~\citep{debugbench} injects bugs using LLMs and relies on online evaluation, which may not reflect realistic software defects. HumanEvalFix~\citep{humanevalfix} contains 164 tasks with human-injected bugs, better capturing real-world error patterns. We therefore adopt HumanEvalFix as our primary benchmark for Python code repair. In hardware, HLSdebugger~\citep{hlsdebugger} generates bugs with LLMs, but its data is not publicly available. RTLFixer~\citep{rtlfixer} collects buggy Verilog programs from LLM-generated incorrect solutions, but these often fail to retain substantial correct logic, limiting the study of precise repairs. We thus build our Verilog benchmark on QiMeng-CodeV-R1~\citep{codev-r1}, which provides high-quality reference implementations and systematic verification.

\paragraph{LLMs for Code Repair.}
Prior LLM-based code repair approaches either use multi-stage pipelines, including error localization, correction, and validation~\citep{agentless,Epperson_2025}, or agent systems with RAG and external tools~\citep{verilogcoder,rtlfixer}. These methods are effective but often slow and costly. Another line of work trains LLMs end-to-end~\citep{humanevalfix,qwen25coder,morepair,slmfix,aligningobjectivellmbasedprogram}, focusing primarily on functional correctness. In contrast, our approach explicitly targets both correctness and repair precision, which is crucial for realistic code repair.

%% file: section/5-conclusion.tex
\section{Conclusion}
\label{sec:conclusion}

In this work, we identify \emph{over-editing} as a fundamental limitation of existing LLM-based code repair approaches that optimize correctness alone. We show that this issue not only increases review burden and harms maintainability, but also weakens error localization and degrades inference efficiency in practical settings. To address this, we propose PRepair, which explicitly encourages minimal yet sufficient edits through self-breaking data generation and the EA-GRPO training objective. Extensive experiments on Python and Verilog Benchmarks demonstrate that PRepair substantially improves repair precision, with $\mathrm{fix}_1@1$ increasing by up to 34.24\% while maintaining stable correctness, and when combined with Speculative Edits, it also accelerates inference, achieving up to 15\% higher decoding throughput highlighting the practical as real-world code assistance.

%% file: appendix/case_study.tex
\section{Case Study}
\label{appendix:case_study}

\subsection{Repair Cases}
\label{repair_cases}
This task requires creating a function that takes a numeric value as a string and returns the nearest integer. When the number is exactly halfway between two integers, the function rounds it away from zero. For example, 14.5 rounds to 15, while -14.5 rounds to -15. The function must correctly handle both positive and negative numbers, as well as numbers with or without decimal points.

The buggy implementation carefully considers string inputs, removes trailing zeros, distinguishes positive and negative .5 values, and applies standard rounding for other numbers. However, it mistakenly rounds positive .5 down and negative .5 up, which is opposite to the intended “round away from zero” behavior.

The baseline GRPO method failed to understand the correct handling in the buggy code and instead rewrote the entire logic, introducing additional errors. Compared to the original buggy implementation, it ignores the careful string-based handling. It mishandles negative .5 values by rounding toward zero, relies on unstable floating-point comparisons, and does not account for empty-string inputs.

In contrast, our method correctly understood the proper handling in the buggy code. It precisely identified the issue of rounding positive .5 down and negative .5 up and made minimal modifications by replacing lines 12 and 14, achieving an accurate and efficient fix.

\begin{tcolorbox}[colback=d_lightred, colframe=d_darkred, enhanced]
\textbf{Buggy Code:}
\begin{lstlisting}
def closest_integer(value):
    from math import floor, ceil

    if value.count('.') == 1:
        # remove trailing zeros
        while (value[-1] == '0'):
            value = value[:-1]

    num = float(value)
    if value[-2:] == '.5':
        if num > 0:
            res = floor(num)
        else:
            res = ceil(num)
    elif len(value) > 0:
        res = int(round(num))
    else:
        res = 0

    return res
\end{lstlisting}
\end{tcolorbox}

\begin{tcolorbox}[colback=d_lightred, colframe=d_darkred, enhanced]
\textbf{Baseline GRPO Results:}
\begin{lstlisting}
import math

def closest_integer(value):
    num = float(value)
    if num - math.floor(num) == 0.5:
        return math.ceil(num)
    elif math.ceil(num) - num == 0.5:
        return math.floor(num)
    else:
        return round(num)
\end{lstlisting}
\end{tcolorbox}

\begin{tcolorbox}[colback=d_lightgreen, colframe=d_darkgreen, enhanced]
\textbf{EA-GRPO Results:}
\begin{lstlisting}[]
def closest_integer(value):
    from math import floor, ceil
    
    if value.count('.') == 1:
        # remove trailing zeros
        while value[-1] == '0':
            value = value[:-1]

    num = float(value)
    if value[-2:] == '.5':
        if num > 0:
            res = ceil(num)
        else:
            res = floor(num)
    elif len(value) > 0:
        res = int(round(num))
    else:
        res = 0

    return res
\end{lstlisting}
\end{tcolorbox}

\subsection{Comparison of Attention Score Heat Map}

\begin{figure*}[!h]
  \includegraphics[width=\linewidth]{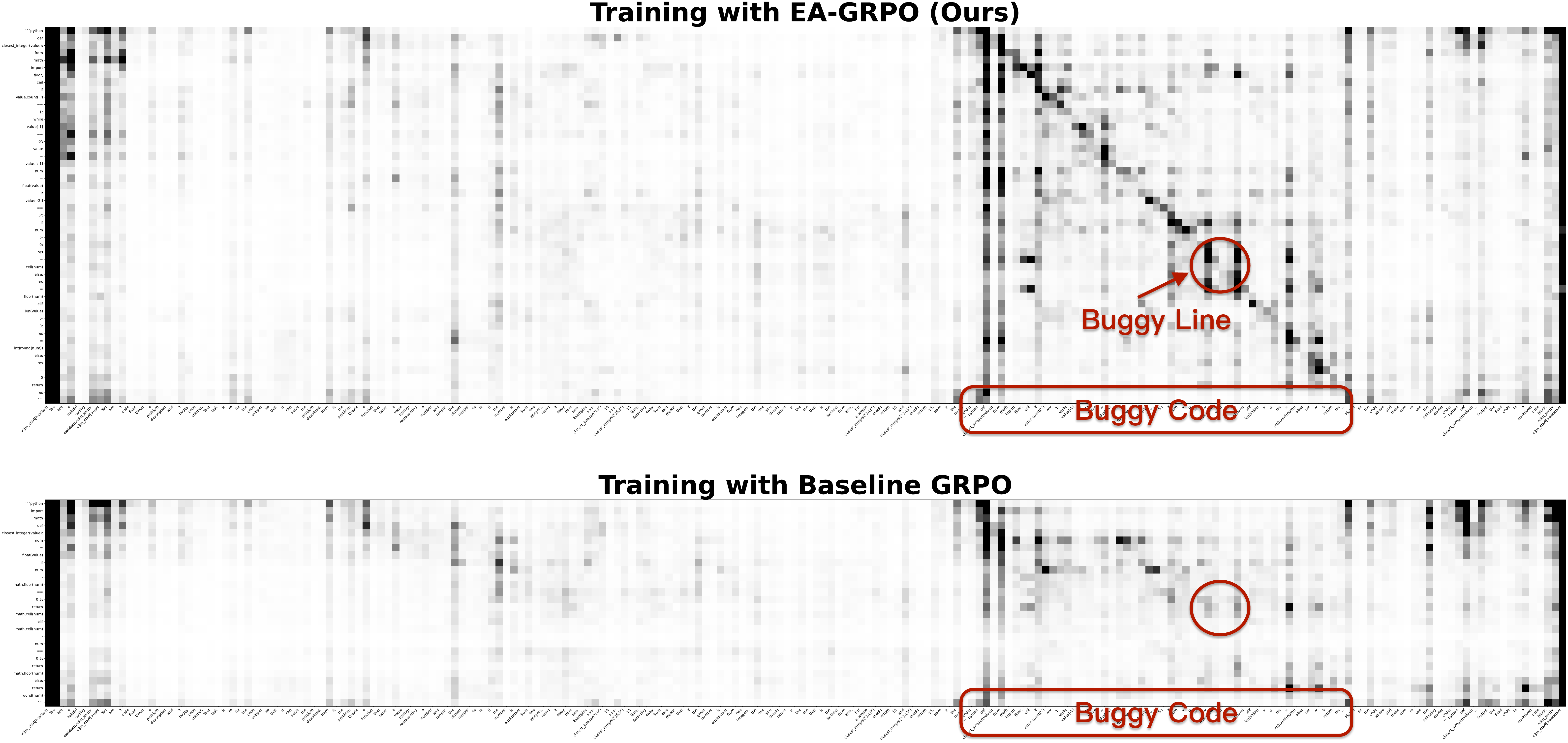}
  \caption {\textbf{Comparison of attention scores in code repair.} The top figure shows the PRepair model trained with EA-GRPO, and the bottom figure shows the model trained with GRPO using correctness-only rewards. The vertical axis corresponds to output tokens, the horizontal axis corresponds to input tokens, and the color intensity indicates the relative magnitude of the attention score.}
  \label{fig:heat_map}
\end{figure*}

To analyze how models specifically attend to repairing the input buggy code, we compute a word-level attention score matrix from the model's token-level attention, using the example in Appendix~\ref{repair_cases}.

Let the input prompt tokens be \(x = \{x_1, \dots, x_n\}\) and the generated output tokens be \(y = \{y_1, \dots, y_m\}\). Denote the model's token-level attention from output to input at layer \(l\) as \(A \in \mathbb{R}^{m \times n}\), where \(A_{ij}\) represents how much output token \(y_i\) attends to input token \(x_j\).

Since tokens may correspond to multiple subword pieces, we first group tokens into words. Let \(M^{\text{in}} \in \mathbb{R}^{n \times N}\) be the input token-to-word mapping, where \(N\) is the number of input words, and \(M^{\text{out}} \in \mathbb{R}^{m \times M}\) the output token-to-word mapping for \(M\) output words. Each entry is normalized by the number of tokens in the corresponding word. Then the word-level attention matrix \(W \in \mathbb{R}^{M \times N}\) is computed as:

\[
W = (M^{\text{out}})^\top \, A \, M^{\text{in}}
\]

Here, \(W_{ij}\) represents how strongly output word \(i\) attends to input word \(j\). Extreme values are clipped at the 98th percentile to improve visualization contrast.

Using this method, we compute and visualize attention matrices for two models: 
\begin{enumerate}
    \item \textbf{Ours}: trained with EA-GRPO.
    \item \textbf{Baseline}: trained with GRPO.
\end{enumerate}

For comparison, the heatmaps in Figure~\ref{fig:heat_map} are plotted vertically, with output words on the vertical axis, input words on the horizontal axis, and color intensity representing the attention scores.

%% file: appendix/details.tex
\section{Implementation Details}
\label{appendix:details}

\subsection{Training Setup}

All RL training experiments are conducted on 8 A100-80GB SXM GPUs for the 7B model and on 8 L40S-48GB GPUs for the 3B model. The training hyperparameters are summarized in Table~\ref{tab:grpo_params}.

\begin{table*}[t]
\centering
\begin{tabular}{l l c l c}
\toprule
\textbf{Category} & \textbf{Parameter} & \textbf{Value} & \textbf{Parameter} & \textbf{Value} \\
\midrule
Algorithm
& Advantage Estimator & GRPO
& Normalize Advantage & True \\
& Use KL in Reward & False
& KL Penalty Type & fixed \\
& KL Coefficient & 0.001
& Target KL & 0.1 \\
\midrule
Policy Optimization
& Learning Rate & $1\times10^{-6}$
& PPO Epochs & 1 \\
& Clip Ratio & 0.2
& Loss Aggregation & token-mean \\
& Entropy Coefficient & 0.0
& Use KL Loss & True \\
& KL Loss Coefficient & 0.001
& KL Loss Type & low\_var\_kl \\
\midrule
Batch \& Token Control
& Train Batch Size & 64
& PPO Mini-batch Size & 64 \\
& PPO Micro-batch / GPU & 2
& Max Tokens / GPU & 16384 \\
\midrule
Rollout Configuration
& Rollout Engine & vLLM
& Rollout Samples ($N$) & 8 \\
& Temperature & 1.0
& Top-$p$ & 1.0 \\
& Top-$k$ & $-1$
& Prompt Length & 2048 \\
& Response Length & 1024
& Sampling Mode & stochastic \\
\midrule
Length Control
& Filter Overlong Prompts & True
& Truncation Strategy & error \\
\midrule
Distributed Training
& Number of Nodes & 1
& GPUs per Node & 8 \\
\bottomrule
\end{tabular}
\caption{\textbf{RL Parameter Setting.} For both the correctness-only reward setting and our PRepair method, we use the same RL hyperparameters to ensure a fair comparison.}
\label{tab:grpo_params}
\end{table*}

\subsection{Inference \& Evaluation}

To reduce statistical bias, we adopt the unbiased estimation method described in Section~\ref{section:metric_design}. We set $n=20$ during evaluation to compute $(\cdot)@1$, $(\cdot)@5$, and $(\cdot)@10$.

\subsubsection{Inference parameters}

For local models, we perform inference using vLLM, with the inference hyperparameters summarized in Table~\ref{tab:inference_params}.

\begin{table}[H]
\centering
\small
\begin{tabular}{cccccc}
\toprule
Max tokens & Temperature & Top-$p$ & Top-$k$ & Min-$p$ \\
\midrule
2048 & 0.7 & 0.8 & 20 & 0 \\
\bottomrule
\end{tabular}
\caption{Inference sampling parameters used for local models.}
\label{tab:inference_params}
\end{table}

\subsubsection{Robust Evaluation for Edit Cost}

Some models may introduce additional comments or reformat the code during repair, which can significantly inflate the measured edit cost and lead to unstable and unfair evaluation. To mitigate this issue, for Python code, we parse the programs into AST and remove all comments as well as redundant whitespace and line breaks before computing the edit cost. Similarly, for Verilog, we use iverilog\footnote{https://github.com/steveicarus/iverilog} to obtain an AST-based representation and eliminate non-semantic characters. This preprocessing ensures that the edit cost reflects only semantic code changes, leading to a fair and consistent evaluation across models.

\subsection{Statistics of Benchmarks}

We summarize the bug types in the two benchmarks in Table~\ref{tab:benchmark_stats}. The results show that the benchmarks cover a wide range of bug categories and subtypes, including diverse logical errors commonly observed in real-world programs.

\begin{table}[H]
\centering
\resizebox{\linewidth}{!}{
\begin{tabular}{lllc}
\toprule
\textbf{Language} & \textbf{Bug Category} & \textbf{Subtype} & \textbf{Count} \\
\midrule
\multirow{6}{*}{Python}
& \multirow{2}{*}{Missing Logic}
& Missing logic & 33 \\
& & Excess logic & 31 \\
& \multirow{2}{*}{O/V Misuse}
& Value misuse & 44 \\
& & Operator misuse & 25 \\
& \multirow{2}{*}{Wrong Logic}
& Variable misuse & 23 \\
& & Function misuse & 8 \\
\cmidrule(lr){2-4}
& \multicolumn{2}{r}{\textbf{Total}} & \textbf{164} \\
\midrule
\multirow{11}{*}{Verilog}
& \multirow{5}{*}{Data-related}
& Bitwise error & 54 \\
& & Value error & 73 \\
& & Width error & 137 \\
& & Arithmetic error & 51 \\
& & Data error & 5 \\
\cmidrule(lr){2-4}
& \multirow{5}{*}{Control-related}
& Comparison error & 12 \\
& & Assignment error & 9 \\
& & Sensitivity list error & 3 \\
& & State error & 4 \\
& & Condition error & 4 \\
\cmidrule(lr){2-4}
& \multicolumn{2}{r}{\textbf{Total}} & \textbf{352} \\
\bottomrule
\end{tabular}
}
\caption{Statistics of bug types in the Python and Verilog code repair benchmarks.}
\label{tab:benchmark_stats}
\end{table}

%% file: appendix/additional_results.tex
\section{Token-Level vs.\ Line-Level Edit Distance}
\label{app:token_vs_line}

\begin{table*}[t!]
\centering
\begin{tabular}{lcccccc}
\toprule
\multirow{2}{*}{Method}
& \multicolumn{3}{c}{\textbf{line-level}}
& \multicolumn{3}{c}{\textbf{token-level}} \\
\cmidrule(lr){2-4} \cmidrule(lr){5-7}
& $\mathrm{fix}_{1}\!@\!1$ & $\mathrm{fix}_{1.5}\!@\!1$ & $\mathrm{fix}_{2}\!@\!1$
& $\mathrm{fix}_{1}\!@\!1$ & $\mathrm{fix}_{1.5}\!@\!1$ & $\mathrm{fix}_{2}\!@\!1$ \\
\midrule
\textbf{GPT4}                       & 2.30  & 3.84  & 5.77  & 4.83  & 7.95  & 13.18 \\
+ Prompt                            & 22.13 & 28.04 & 33.92 & 24.94 & 33.15 & 38.01 \\
\cmidrule(lr){1-7}
\textbf{Gemini2.0-flash}            & 19.06 & 24.01 & 30.57 & 19.72 & 31.28 & 35.34 \\
+ Prompt                            & 42.33 & 48.44 & 53.49 & 45.14 & 54.26 & 57.27 \\
\cmidrule(lr){1-7}
\textbf{Qwen2.5-Coder-7B}           & 36.70 & 42.86 & 48.98 & 43.49 & 47.23 & 49.69 \\
+ Prompt                            & \underline{38.00} & \underline{43.81} & \underline{49.59} & \underline{44.19} & \underline{48.14} & \underline{50.30} \\
+ GRPO                              & 8.49  & 12.93 & 23.85 & 16.34 & 30.80 & 38.42 \\
\rowcolor{blue!10} \cellcolor{white} + EA-GRPO
                                    & \textbf{68.11} & \textbf{68.11} & \textbf{68.59} & \textbf{67.66} & \textbf{68.39} & \textbf{68.59} \\
\bottomrule
\end{tabular}
\caption{\textbf{Line-level vs.\ token-level edit distance on Verilog.} We report $\mathrm{fix}_p@1$ with $p \in \{1, 1.5, 2\}$ under both granularities. \textbf{Bold} indicates the best result, and \underline{underline} indicates the second best in the same model. The ranking of methods is fully consistent across the two granularities.}
\label{tab:token_vs_line}
\end{table*}

Our $\mathrm{fix}_p@k$ metric is built on line-level edit distance. This choice is deliberate and task-aligned, and we further explore a token-level variant to verify that our conclusions are robust to the granularity of the edit cost.

\paragraph{Semantic consistency.} Compared with the commonly used token-level edit distance, line-based edit distance better preserves consistency in semantic importance. Token-level edit distance is often too fine-grained and can underestimate semantic changes. For example, replacing \texttt{a = b} with \texttt{a = c} changes only one token out of three at the token level, yet this modification completely alters the assignment semantics. At the line level, the edit cost is one full line, which more faithfully reflects the actual impact of the change.

\paragraph{Alignment with real-world development.} A central motivation of our work is to reduce developers' review burden. In real workflows, code changes are inspected at the line level: tools such as \texttt{git diff} and Unix \texttt{diff} report modifications line by line, and code review is conducted line by line. Developers do not review code at the token or AST level. Line-based edit cost is therefore more consistent with practical usage scenarios.

\paragraph{Empirical comparison.}
We additionally evaluate Verilog baselines under a token-level version of $\mathrm{fix}_p@1$. As shown in Table~\ref{tab:token_vs_line}, the performance trends under token-level and line-level metrics are fully consistent: EA-GRPO remains the best method by a large margin under both granularities, while vanilla GRPO remains the weakest on the fix metric.

%% file: appendix/spec_edit.tex
\section{Speculative Edits}
\label{spec_edit}

\begin{table}[t]
\centering
\small
\resizebox{\columnwidth}{!}{
\begin{tabular}{llrrrr}
\toprule
\textbf{Lang} & \textbf{Method} & \textbf{TPS} & \textbf{Draft} & \textbf{Acc.} & \textbf{AR} \\
\midrule
Verilog & Origin   & 24.34 & 229,650 & 114,467 & 0.498 \\
        & EA-GRPO  & \textbf{28.15} & 214,404 & 128,752 & \textbf{0.601} \\
        & GRPO     & 15.87 & 295,744 & 76,786  & 0.260 \\
\midrule
Python  & Origin   & 29.70 & 13,404  & 7,735  & 0.577 \\
        & EA-GRPO  & \textbf{31.51} & 13,065 & 8,847 & \textbf{0.677} \\
        & GRPO     & 28.22 & 14,105 & 6,751 & 0.479 \\
\bottomrule
\end{tabular}}
\caption{Decoding performance with N-gram speculative decoding.
TPS denotes throughput (tokens/s), Acc. denotes accepted tokens, and AR denotes acceptance rate.}
\label{table:spec_edit}
\end{table}

To analyze the acceleration benefits of our method, we provide an analytical approximation that relates the program repair objective to the efficiency of Prompt Lookup Decoding under conservative assumptions. Prompt Lookup Decoding retrieves N-gram matches from the prompt at each decoding step and use them as draft tokens.

\subsection{Acceptance Derivation}

Let a buggy program be represented as a sequence of lines \(X = \{x_1, x_2, \dots, x_n\}\), where \(n = |X|\) denotes the total number of lines. The repair process produces a corrected program \(Y = \{y_1, y_2, \dots, y_m\}\). Our EA-GRPO objective explicitly minimizes the normalized Distance Edit Cost, denoted as \(\mathbf{D}_\mathrm{EC}(X, Y)\), which measures the fraction of modified lines between \(X\) and \(Y\).

For a given program \(X\), the expected number of modified lines \(M\) is approximated as:
\begin{equation}
M = |X| \cdot \mathbf{D}_\mathrm{EC}(X, Y).
\end{equation}

In N-gram speculative decoding, draft tokens are obtained by performing an N-gram lookup over the input prompt (the buggy code \(X\)). For analytical tractability, we adopt a conservative approximation where draft tokens are aligned and verified at the line level: a line contributes to successful speculative acceptance only if it remains unchanged in the repaired output. Under this assumption, the probability that a randomly selected line is accepted, denoted as \(R_{\text{line}}\), is given by:
\begin{equation}
R_{\text{line}} = \frac{|X| - M}{|X|} = 1 - \mathbf{D}_\mathrm{EC}(X, Y).
\end{equation}

Although speculative decoding operates at the token level, this approximation captures the dominant behavior in code repair, where edits typically disrupt token continuity within modified lines. Therefore, we approximate the token-level acceptance rate \(R\) by the line-level acceptance ratio:
\begin{equation}
R \approx R_{\text{line}} = 1 - \mathbf{D}_\mathrm{EC}(X, Y).
\end{equation}

This relation indicates that the speculative acceptance rate is inversely correlated with the edit cost. By explicitly minimizing \(\mathbf{D}_\mathrm{EC}(X, Y)\), EA-GRPO effectively increases \(R\), transforming the input buggy program into a high-fidelity implicit draft for speculative decoding.

\subsection{Throughput Derivation}

Given a speculative window of \(K\) draft tokens, we analyze the expected number of tokens generated per target model verification step. Let the random variable \(X\) denote the number of tokens accepted before the first mismatch, where \(X \in \{1, 2, \dots, K+1\}\). Specifically, \(X = i+1\) if the first \(i\) draft tokens are accepted and the \((i+1)\)-th token is rejected, except for the case where all \(K\) draft tokens are accepted.

Under the assumption that each draft token is independently accepted with probability \(R\), the probability mass function is:
\begin{equation}
P(X = i+1) =
\begin{cases}
R^i (1 - R), & 0 \le i < K, \\
R^K, & i = K.
\end{cases}
\end{equation}

The expected number of tokens produced per verification step is:
\begin{align*}
&\mathbb{E}[X] \\
&= \sum_{i=0}^{K-1} (i+1)\,R^i(1-R) + (K+1)R^K \\
&= (1-R)\sum_{i=0}^{K-1}(i+1)R^i + (K+1)R^K \\
&= (1-R)\sum_{j=1}^{K} j R^{\,j-1} + (K+1)R^K \\
&= (1-R)\,\frac{d}{dR}\!\left(\sum_{j=0}^{K} R^{\,j}\right) + (K+1)R^K \\
&= (1-R)\,\frac{d}{dR}\!\left(\frac{1-R^{K+1}}{1-R}\right) + (K+1)R^K \\
&= (1-R)\,\frac{-(K+1)R^K(1-R) + (1-R^{K+1})}{(1-R)^2} \\
&+ (K+1)R^K \\
&= \frac{1 - (K+1)R^K + K R^{K+1}}{1-R} + (K+1)R^K \\
&= \frac{1 - R^{K+1}}{1-R}.
\end{align*}

Substituting the approximation \(R \approx 1 - \mathbf{D}_\mathrm{EC}(X, Y)\), we obtain:
\begin{equation}
E \approx \frac{1 - (1 - \mathbf{D}_\mathrm{EC}(X, Y))^{K+1}}{\mathbf{D}_\mathrm{EC}(X, Y)}.
\end{equation}

Since the N-gram lookup latency is negligible compared to the target model verification cost, the system throughput (measured as tokens per second) scales proportionally with \(E\). Relative to the baseline decoding scheme where \(E = 1\), the throughput improvement factor is therefore approximately:
\begin{equation}
T \propto \frac{1 - (1 - \mathbf{D}_\mathrm{EC})^{K+1}}{\mathbf{D}_\mathrm{EC}}.
\end{equation}
We consider the throughput function
\begin{equation}
T \propto f(D) = \frac{1 - (1-D)^{K+1}}{D}, \quad D \in (0,1).
\end{equation}
Taking the derivative with respect to \(D\) gives
\begin{align*}
&f'(D)=\\ 
&\frac{D (K+1)(1-D)^K - \big(1 - (1-D)^{K+1}\big)}{D^2}.
\end{align*}
The numerator can be simplified as
\begin{align*}
g(D) = (K+1) D (1-D)^K & - 1 + (1-D)^{K+1} \\
& < 0, \quad \forall D \in (0,1),
\end{align*}
which implies \(f'(D) < 0\). Therefore, \(f(D)\) is strictly decreasing with \(D\), i.e.,
\begin{equation}
\text{as } \mathbf{D}_\mathrm{EC} \text{ decreases, } T \text{ increases.}
\end{equation}

This analysis shows that as EA-GRPO reduces the edit cost, the system transitions into a high-efficiency regime where the expected token yield grows non-linearly with decreasing \(\mathbf{D}_\mathrm{EC}\). This theoretical trend is consistent with our empirical observations in Figure \ref{fig:spec_edit} and Table~\ref{table:spec_edit}, where EA-GRPO significantly improves both acceptance rate and end-to-end decoding throughput.

%% file: appendix/grpo.tex
\section{Preliminary of GRPO}
\label{detals_grpo}
Group Relative Policy Optimization (GRPO) is an on-policy reinforcement learning algorithm built upon the Proximal Policy Optimization (PPO) framework. GRPO removes the value model to significantly reduce inference cost, while introducing group relative advantage estimation to more accurately assess the quality of model outputs. Furthermore, a KL-divergence penalty is incorporated to stabilize policy updates and prevent the policy from deviating excessively against the reference model.

Given a group $\mathcal{G}$ with rewards
$\{\mathcal{R}^{\mathcal{G}}_i\}_{i \in \mathcal{G}}$,
the group-normalized advantage is computed as
\[
\mathcal{A}^{\mathcal{G}}_i
=
\frac{
\mathcal{R}^{\mathcal{G}}_i
-
\mathrm{mean}\big(\mathcal{R}^{\mathcal{G}}_j\big)
}{
\mathrm{std}\big(\mathcal{R}^{\mathcal{G}}_j\big)
}
\]

The computed advantage is broadcast to all tokens of the corresponding output.
Model parameters are updated using the GRPO objective with a KL divergence constraint:
\[
\begin{aligned}
& \mathcal{J}(\theta)
=
\mathbb{E}\Bigg[
\frac{1}{|\mathcal{G}|}
\sum_{i \in \mathcal{G}}
\frac{1}{|o_i|}
\sum_{t=1}^{|o_i|}
\min\Big(
r_{i,t}(\theta)\mathcal{A}^{\mathcal{G}}_i, \\
& \mathrm{clip}\big(r_{i,t}(\theta), 1-\epsilon, 1+\epsilon\big)
\mathcal{A}^{\mathcal{G}}_i
\Big)
-
\gamma
\mathrm{KL}\!\left(
\pi_{\theta}
\|
\pi_{\theta_{\mathrm{old}}}
\right)
\Bigg]
\end{aligned}
\]
where
\[
r_{i,t}(\theta)
=
\frac{
\pi_{\theta}(o_{i,t} \mid x, o_{i,<t})
}{
\pi_{\theta_{\mathrm{old}}}(o_{i,t} \mid x, o_{i,<t})
}
\]
is the importance sampling ratio at token $t$, and $\gamma$ controls the strength of the KL regularization.

%% file: appendix/prompts.tex
\section{Prompts}

In this section, we detail the prompt utilized in the process of Self-Breaking and Self-Repairing.

The following is the prompt we use for Self-Breaking.

\begin{prompt}
\textbf{Prompt: }
You are a code breaker. 
Your task is to subtly introduce bugs into the provided code while preserving its syntactic correctness and overall structure. \\
You may modify, replace, or delete lines to insert bugs that cause execution failures or incorrect results. Make sure your bugs are not obvious and are challenging to detect and fix. 
First, briefly explain the bugs you introduced. Then, output only the modified code, wrapped in a code block, without any additional comments.\\
**Please keep the code format unchanged and only insert the necessary modifications.**\\
Here is the golden code to break:\\
\lstinline|```{language}|\\
\lstinline|{code}|\\
\lstinline|```|\\
Here is the problem associated with the code:\\
\lstinline|{problem}|
\end{prompt}
\vspace{8pt}

The following is the prompt we use for Self-Repairing.

\begin{prompt}
\textbf{Prompt: }
You are a code fixer. Given a problem description and a buggy code snippet, Your task is to fix the code snippet so that it can solve the problem described. \\

Here is the problem:\\
\lstinline|{problem}|\\
Here is the buggy code:\\
\lstinline|```{language}|\\
\lstinline|{buggy_code}|\\
\lstinline|```|\\
Output the fixed code in a markdown code block.
\end{prompt}
\vspace{8pt}

